%% file: lcsr15.tex
\documentstyle[epsfig,graphicx,12pt]{article}
\renewcommand{\theequation}{\arabic{equation}}
\newcommand{\ba}{\begin{array}}
\newcommand{\ea}{\end{array}}

\newcommand{\be}[1]{
\begin{eqnarray}\label{#1}}
\newcommand{\ee}{\end{eqnarray}}
\newcommand{\insertfig}[2]{\mbox{\epsfxsize=#1cm \epsfbox{#2.eps}}}

\newcommand{\btab}{\begin{tabular}}
\newcommand{\etab}{\end{tabular}}

\newcommand{\ci}[1]{\cite{#1}}

\newcommand{\re}[1]{\ref{#1}}

\newcommand{\edd}{\end{document}}

\newcommand{\alf}{\ifmmode\alpha \else$\alpha \ $\fi}
\newcommand{\bt}{\ifmmode\beta \else$\beta \ $\fi}
\newcommand{\gm}{\ifmmode\gamma \else$\gamma \ $\fi}
\newcommand{\Dl}{\ifmmode\Delta \else$\Delta \ $\fi}
\newcommand{\eps}{\ifmmode\varepsilon \else$\varepsilon \ $\fi}
\newcommand{\dl}{\ifmmode\delta \else$\delta \ $\fi}
\newcommand{\et}{\ifmmode\eta \else$\eta \ $\fi}
\newcommand{\vphi}{\ifmmode\varphi \else$\varphi \ $\fi}
\newcommand{\om}{\ifmmode\omega \else$\omega \ $\fi}
\newcommand{\pl}{\ifmmode\partial \else$\partial \ $\fi}
\newcommand{\ps}{\ifmmode\psi \else$\psi \ $\fi}
\newcommand{\sg}{\ifmmode\sigma \else$\sigma \ $\fi}
\newcommand{\phf}{\ifmmode\varphi^4 \else$\varphi^4 \ $\fi}
\newcommand{\Lam}{\ifmmode\Lambda \else$\Lambda$\fi}

\newcommand{\ppp}[1]{%
        \setbox0=\hbox{#1}%
        \kern-.02em\copy0\kern-\wd0
        \kern+.04em\copy0\kern-\wd0
        \kern-.02em\raise.0217em\box0}

\newcommand{\lsim}{
 \mathrel{\setbox0=\hbox{$<$}\raise0.6ex\copy0\kern-\wd0
 \lower0.65ex\hbox{$\sim$}}}

\newcommand{\gsim}{
 \mathrel{\setbox0=\hbox{$>$}\raise0.6ex\copy0\kern-\wd0
 \lower0.65ex\hbox{$\sim$}}}

\newcommand{\PRD}[3]{Phys.\ Rev.\ D {\bf {#1}}, {#2} ({#3})}

\newcommand{\PLB}[3]{Phys.\ Lett.\ B {\bf {#1}}, {#2} ({#3})}

\begin{document}
\renewcommand{\thefootnote}{\fnsymbol{footnote}}
\makebox[2cm]{}\\[-1in]
\begin{flushright}
\begin{tabular}{l}
\\
TPR-00-12\\
\end{tabular}
\end{flushright}
\vskip0.4cm
\renewcommand{\thefootnote}{\fnsymbol{footnote}}
\begin{center}
{\bf\Large Power corrections to the process 
$\gamma\gamma^*\rightarrow \pi\pi$ \\in the
Light-Cone Sum Rules approach
}

\vspace{0.5cm}

N. Kivel$^{a,b}$, L. Mankiewicz$^{c,d}$

\vspace{0.5cm}

\begin{center}

{\em$^a$Institut f\"ur Theoretische Physik, Universit\"at Regensburg \\
D-93040 Regensburg, Germany }

{\em $^b$Petersburg Nuclear Physics Institute,
  188350, Gatchina, Russia
}

{\em $^c$ N. Copernicus Astronomical Center, ul. Bartycka 18,
PL--00-716 Warsaw, Poland}

{\em $^d$ Andrzej Soltan Institute for Nuclear Studies,
Warsaw, Poland}

\end{center}

\centerline{\bf version from  \today}

 \end{center}
 \vspace{1.5cm}
\begin{abstract}
  We applied QCD Light Cone Sum Rules to estimate power corrections to the
  helicity-conserving amplitude in the process $\gamma^*\gamma\rightarrow
  \pi\pi$. We found that above $Q^2 \sim 4$ GeV$^2$ power corrections are
  numerically small and the twist-2 part dominates.The amplitude can be
  reliably calculated in this region using models of $2 \pi$ distribution
  amplitudes as an input. We found that the magnitude of the NLO corrections
  depends rather strongly on the normalization of the gluonic distribution
  amplitude.
\end{abstract}

\section{\normalsize \bf Introduction}

Hadron production in the reaction $\gamma^*\gamma\rightarrow {\mbox
{hadron(s)}}$ has been a subject of considerable interest for a long
time, both from the experimental \ci{CELLO,CLEO} and theoretical 
\ci{Budnev, Lepage, Efremov}
points of view. The key role in QCD description of such processes
is played by the QCD factorization theorem.  For example, QCD
factorization has been successfully applied to the reaction
$\gamma^*\gamma\rightarrow\pi^0$ \ci{Lepage, Efremov}.  
The $F^{\gamma\pi}(Q^2)$
form-factor data obtained by the CELLO and CLEO collaborations are in a good
agreement with the available QCD analysis, see for example 
\ci{RadRus, Schmed, Jakob}.

Recently it has been proposed \ci{Dittes,Diehl} to investigate a similar
process $\gamma^*\gamma\rightarrow\pi\pi$ when the two pion state has
a small invariant mass. It has been argued that QCD factorization
applies to this case as well \ci{Freund}. The resulting amplitude
depends on new non-perturbative objects, the so-called two-pion
distribution amplitudes ($2\pi$DA's). They are given by matrix
elements of twist-2 QCD string operators between vacuum and the
two-pion state \ci{Diehl, Polyakov}.  Moreover, $2\pi$DA's can be
related by the crossing symmetry to skewed parton distributions
\cite{Ji97,Rad97} which recently have been subject of considerable
interest.

Furthermore, in the recent paper \ci{Diehl1} it has been argued that
experimental studies of 2$\pi$ production cross-section are
possible with existing $e^+e^-$ facilities.

Formally, dominance of the leading-twist amplitude is guaranteed only
at a very large $Q^2$. For the process
$\gamma^*\gamma\rightarrow\pi\pi$ the bulk of the twist-2 amplitude
arises from the handbag diagram. In analogy with the
$\gamma^*\gamma\rightarrow\pi^0$ reaction one expects that the
leading-twist contribution dominates the amplitude even for moderate
values of $Q^2\sim 4-10$ GeV$^2$. For lower values of $Q^2$, the
power-suppressed corrections are certainly important. Note that
preliminary estimates show that most of the
$\gamma^*\gamma\rightarrow\pi\pi$ events which have been seen in the
CLEO data are in the region $Q^2\sim 1-5$ GeV$^2$ \ci{CLEO,Sav}. In this
region a reliable estimate of the amplitude including the
power-suppressed contributions is crucial.

From the theoretical point of view problems encountered in an analysis of
power-suppressed contributions to two-pion and one-pion production
amplitudes are very similar. Terms suppressed as $1/Q^2$ can arise
from different space-time configurations. Production of states with
more than 2 partons by interaction of two electromagnetic currents at small
transverse distances can be accounted for by the standard Operator
Product Expansion (OPE) technique. However, as the second photon is
real, there is yet another configuration which results in a
power-suppressed correction. The real photon can turn into hadrons
long before the interaction with the virtual one. It occurs at large
transverse distances between two electromagnetic currents. Such a
not-factorizable term is known in the literature as the `end-point' or
`soft' contribution.

In this paper we use the light-cone sum rules (LCSR) method to
evaluate the $\gamma^*\gamma\rightarrow\pi\pi$ amplitude, including
the power suppressed contributions. The main advantage of this
technique is that it allows to take into account both factorizable and
non-factorizable corrections. Recently, LCSR were successfully applied
to describe different pion form-factors \ci{Schmed, Khod, Braun99} 
in a $Q^2$ range from 1 to $\sim 10$ GeV$^2$.

This paper is organized as follows. In the following section we present the
definition of the $\gamma^*\gamma\rightarrow\pi\pi$ amplitude and set
up the notation. In section 2 models of $2\pi$ distribution amplitudes are
introduced. Sections 2 and 3 are devoted to discussion of the LCSR to the LO
and NLO accuracy, respectively. In section 4 we present  numerical
analysis. Finally, we summarize. 
The appendix contains definitions of the NLO Wilson
coefficient functions.
 
\section{\normalsize \bf General definitions}

Kinematics of the reaction $\gamma^*(q) \gamma(q') \to \pi(k_1) \pi(k_2)$
can conveniently  be described in terms of a pair of light-like vectors $p,\ z$
which obey
\be{nvectors}
\ba{l}
p^2 = z^{2} = 0, \, p \cdot z \not = 0
\ea
\ee
and define longitudinal directions. Here $p \cdot z = p_\mu z^\mu$. 
Let $P$ and $k$ denote total and relative
momenta of the $\pi$ meson pair, respectively,
\be{pivectors}
\ba{l}
P^2=(k_1+k_2)^2=W^2, k^2=(k_1-k_2)^2=4 \, m_\pi^2 - W^2, P\cdot k = 0 
\ea
\ee

The initial and final states momenta can be decomposed as

\be{lcexpansion}
\ba{l}
\displaystyle
q = p  - \frac{Q^2}{2 (p\cdot z)}z, \quad q^2 = - Q^2\, 
\hspace*{8mm}
q^\prime = \frac{Q^2+W^2}{2 (p\cdot z)}\, z, \quad q^{\prime \, 2} = 0
\\[4mm]\displaystyle
P = q + q^\prime = p + \frac{W^2}{2 (p\cdot z)} \, z, \quad  P^2 = W^2
\\[4mm] \displaystyle
k = \xi p - \frac{\xi W^2}{2 (p\cdot z)} \, z  + k_\perp
\ea
\ee
The longitudinal momentum distribution between pions is described by the
variable $\xi = (k\cdot z)/(p\cdot z)$. Alternatively,
$$
\xi = \beta \cos\theta_{\rm cm}\, ,
$$
where $\theta_{\rm cm}$ is the polar
angle of the pion momentum in the CM frame with respect to
the direction of the total momentum $P$ and $\beta$ is the velocity of
produced pions in the center-of-mass frame
\begin{eqnarray}
\nonumber
\beta=\sqrt{1-\frac{4 m_\pi^2}{W^2}}\, .
\end{eqnarray}

The amplitude of
hard photo-production of two pions
is defined by the following matrix element
between vacuum and two pions state:
\be{defT}
T^{\mu\nu} = i \int d^4x e^{-ix\cdot {\bar q}}
\langle 2 \pi (P, k)| T J^\mu(x/2) J^\nu(-x/2) | 0 \rangle\, ,
\mskip10mu {\bar q}=\frac12(q-q')\, 
\ee
where $J^\mu(x)$ denotes quark electromagnetic current.  Hard
photo-production corresponds to the limit
$Q^2 \gg \,  W^2 \geq \Lambda_{\rm QCD}^2$
where the amplitude (\ref{defT}) can be represented as an expansion
in terms of powers of $1/Q$. According to the factorization theorem
the leading twist term in the expansion can be written as a convolution of
hard  and soft blocks. The coefficient functions can be calculated from 
appropriate partonic subprocesses $\gamma^*+\gamma\rightarrow 
\bar{q}+q$ or $\gamma^*+\gamma\rightarrow g+g$. 

According to the analysis of Ref. \ci{KMP} to the leading twist accuracy the 
amplitude $T^{\mu\nu}$ is a sum of two
terms
\be{Tdecompos}
T^{\mu\nu}(q,q',P,k)=\frac i2(-g^{\mu\nu})_T T^{\gamma\pi\pi}_0(q,q',P,k)+
\frac i2\frac{k_\perp^{(\mu} k_\perp^{\nu)}}{W^2} \,T^{\gamma\pi\pi}_2(q,q',P,k)
\ee
where $(-g^{\mu\nu})_T=\left( \frac{p^\mu z^{\nu} + p^\nu z^{\mu}}{p \cdot z}
  - g^{\mu\nu}\right)$ is the metric tensor in the transverse space and
$k_\perp^{(\mu} k_\perp^{\nu)}$ denotes traceless, symmetric tensor product of
relative transverse momenta (\re{kinematics}).

The leading-order, leading-twist amplitude $T_0^{\gamma\pi\pi}$ describes
scattering of two photons with equal helicities, related by crossing to the
photon helicity-conserving DVCS on a pion.  At the NLO there is a new
contribution $T_2^{\gamma\pi\pi}$ from collisions of photons with opposite
helicities, related to photon helicity-flip contribution to DVCS \ci{JiH}. 
In terms of the Operator Product Expansion the latter amplitude singles out
 a twist-2
tensor gluon operator which cannot be studied in deep-inelastic scattering
(DIS) on a pion (or nucleon) target \ci{KMP}.

Note that as the amplitude $T^{\mu\nu}$ is dimensionless, twist-2 amplitude
$T_0^{\gamma\pi\pi}$ depends on $Q^2$ only logarithmically through the running
coupling and QCD evolution effects. To see this it is convenient to develop an
appropriate power counting in an infinite momentum frame. For definiteness we
assume that the pion pair is moving in the positive ${\hat z}$ direction and
$p^+$ and $z^-$ are the only nonzero component of $p$ and $z$, respectively.
Then the infinite momentum frame can be understood as $p^+ \sim Q \rightarrow
\infty$ with a fixed $(p\cdot z)\sim 1$. From (\re{lcexpansion}) it 
follows that in
this frame $(k\cdot z)\sim 1$ and $k_\perp \sim Q^0$. This determines the power
counting in $Q$ for all twist-2 amplitudes and from (\re{Tdecompos}) we find
that $T_0^{\gamma\pi\pi}$ is $O(1)$ as far as powers of $Q$ are concerned.

In this paper we consider power corrections to the amplitude
$T^{\gamma\pi\pi}_0$ only. It is expected to be the most important term
numerically as it appears already at the Born level.  To the NLO accuracy one
has \ci{KMP}:
\be{NLOLz0}
\ba{l}
\displaystyle
T^{\gamma\pi\pi}_0=T^{pert}_0
=
\left( \sum e_q^2 \right) \int_0^1 du\,
\Phi^Q(u,\xi,W^2) \left[C^0_q(u)+
{\textstyle \frac{\alf_S(Q^2)}{4\pi}}
C^1_q(u)\right]
\\[8mm]
\displaystyle
\hspace*{10mm}
- \left(\sum e_q^2\right)
\int_0^1 du\,
\Phi^G(u,\xi,W^2)\left[
{\textstyle \frac{\alf_S(Q^2)}{4\pi}} C^1_g(u)
\right] \, .
\ea
\ee
Coefficient functions $C^0_q,C^1_q, C^1_g$ can be found in \ci{KMP}.  The
quark and gluon $2\pi$DA's are defined as matrix elements of the light-cone
string operators:
\be{quarkDA}
\langle \pi\pi(P, k) | \frac{1}{N_f}\sum_q {\bar q}(z) {\hat z} q(-z) |0\rangle
= (p\cdot z) \int_0^1 du \, \Phi^Q(u,\xi,W^2) \, e^{i(2u-1)(p\cdot z)} \, ,
\ee
\be{gluonDA}
\langle \pi\pi(P,k) | z^\mu z^\nu \,
G^{\mu}_{\phantom{\mu}\alpha}(z) G^{\alpha\nu}(-z)
|0\rangle
=( p\cdot z)^2\int_0^1 du\ 
\Phi^G (u, \xi, W^2 )\ e^{i (2u-1) (p\cdot z)}\; ,
\ee
Note that distribution amplitudes $\Phi^Q(u,\xi,W^2)$ and $\Phi^G (u, \xi, W^2
)$ depend also on a factorization scale $\mu$.

\section{\normalsize \bf Models for 2-pion distribution amplitudes }

In this section we describe briefly the main properties of the distribution
amplitudes introduced in (\re{quarkDA}) and (\re{gluonDA}) and discuss
a model which has been used to obtain estimates for the magnitude of
power-suppressed corrections. 

Due to the positive $C$-parity of the pion pair, $2\pi$ distribution
amplitudes have the following symmetry properties
\be{Cparity}
\ba{l}\displaystyle
\Phi^Q(u,\xi,W^2)=-\Phi^Q(1-u,\xi,W^2)= \Phi^Q(u,-\xi,W^2),
\\[4mm]\displaystyle
\Phi^G(u,\xi,W^2)=\Phi^G(1-u,\xi,W^2)=\Phi^G(u,-\xi,W^2).
\ea
\ee

The factorization scale dependence is governed by the ERBL evolution
equations \cite{ER78,BL79}. As it is well known, in the leading logarithmic
approximation their solution has a form of an expansion in terms of
Gegenbauer polynomials:
\be{razhlq}
\Phi^Q(u,\xi, W^2|\, \mu )=6u(1-u)
\sum_{\scriptstyle n=1 \atop \scriptstyle {\rm odd}}^{\infty}
 B_{n}(\xi,W^2|\, \mu) \, C_n^{3/2}(2 u-1),
\ee
\be{razhlg}
\Phi^G(u,\xi, W^2|\, \mu )=
30\ u^2(1-u)^2
\sum_{\scriptstyle n=0 \atop \scriptstyle {\rm even}}^\infty
A_{n}(\xi,W^2 |\, \mu)\,  C^{5/2}_n(2 u-1) ,
\ee
Coefficients $B_n$
and $A_n$ mix under evolution. 

In the next step one expands both distribution amplitudes in the partial waves
of the final 2-pion system \ci{Polyakov}. As a result one introduces an
expansion of $B_{n}(\xi,W^2 |\, \mu)$ and $A_n(\xi,W^2 |\, \mu)$ in terms of
Legendre polynomials $P_l(\xi)$ \ci{Polyakov, KMP}:
\be{ABexp}
\ba{l}\displaystyle
B_{n}(\xi,W^2 |\, \mu)=
\sum_{\scriptstyle l=0 \atop \scriptstyle {\rm even}}^{n+1} 
B_{nl}(W^2 |\, \mu)P_l(\xi)
\\[4mm]\displaystyle
A_n(\xi,W^2 |\, \mu)=\sum_{\scriptstyle l=0 \atop \scriptstyle {\rm
    even}}^{n+2} 
A^G_{nl}(W^2 |\, \mu)P_l(\xi)
\ea  
\ee

Additional constraints on $2 \pi$ distribution amplitudes are provided by soft
pion theorems,  
see \ci{Polyakov}:
\be{softT}
\ba{l}\displaystyle
\Phi^Q(u,\xi=1,W^2=0)=\Phi^Q(u,\xi=-1,W^2=0)=0.
\\[4mm]\displaystyle
\Phi^G(u,\xi=1,W^2=0)=\Phi^G(u,\xi=-1,W^2=0)=0.
\ea
\ee
Finally, crossing symmetry allows to relate moments of distribution amplitudes
to forward matrix elements of twist-2 operators which determine moments of
pion quark and gluon structure functions, see \ci{Polyakov} for details.
One finds, e.g.
\be{norm}
\ba{l}\displaystyle
\int_0^1 du (2u-1)\, \Phi^Q(u,\xi=1,W^2=0)=-\frac1{N_f}M^Q (1-\xi^2)
\\[4mm]\displaystyle
\int_0^1 du\, \Phi^G(u,\xi=1,W^2=0)=-\frac12M^G(1-\xi^2)
\ea
\ee
where $M^Q$ and $M^G$ are momentum fractions carried by quarks and 
gluons in a pion:
\be{defM}
\ba{l}\displaystyle
M^Q(\mu) =\int_0^1 du u\sum_q (q_{\pi}(u,\mu)+\bar q_{\pi}(u,\mu))
\\[4mm]\displaystyle
M^G(\mu) =\int_0^1 du ug_{\pi}(u,\mu) \, .
\ea
\ee

At asymptotically large $\mu^2\rightarrow \infty$ only the lowest terms in
(\re{razhlq}), (\re{razhlg}) contribute. Combining constrains (\re{norm}) with
(\re{ABexp}) one easily finds:
\be{DAas}
\ba{l}
\displaystyle
\Phi^G_{as} (u, \zeta, W^2=0 )=
- 15 u^2(1-u)^2 M^G_{as} (1-\xi^2)\, ,
\\[4mm]  \displaystyle
\Phi^Q_{as} (u, \zeta, W^2=0)=
-30 u(1-u)\ (2u-1)\frac1{N_f} M^Q_{as} (1-\xi^2)  \, .
\ea  
\ee
where
\be{Masymp}
M^Q_{as}=\frac{N_f}{N_f+4C_F},\quad  M^G_{as}=\frac{4C_F}{N_f+4C_F} \, .
\ee
For $W^2 \not= 0$ one writes 
\be{DAasW}
\ba{l}
\displaystyle
\Phi^G_{as} (u, \xi, W^2 )=
- 15 u^2(1-u)^2 M^G_{as} B(\xi,W^2)\, ,
\\[4mm]  \displaystyle
\Phi^Q_{as} (u, \xi, W^2)=
-30 u(1-u)\ (2u-1)\frac1{N_f} M^Q_{as} B(\xi,W^2) \, .
\ea 
\ee
Function $B(\xi,W^2)$ is related to coefficients
$B_1$ and $A_0$ from (\re{ABexp})
\be{defB}
B(\xi,W^2)=-\frac{2}{M^G_{as}}A_0(\xi,W^2|\mu^2=\infty)=
-\frac35\frac{N_f}{M^Q_{as}}B_1(\xi,W^2|\mu^2=\infty).
\ee
In the limit $W^2\rightarrow 0$ one finds from (\ref{norm}) 
that $B(\xi,W^2=0)=(1-\xi^2)$.

In numerical calculations presented in this paper we have used a model which
retains simple analytical form of asymptotic distributions amplitudes
(\re{DAasW}), but incorporates nontrivial information about pion structure at
a scale of the order of a few GeV$^2$ \ci{Diehl1}. Assuming dominance of the lowest
conformal wave one finds
\be{DAmodel}
\ba{l}
\displaystyle
\Phi^G (u, \xi, W^2 )=
- 15 u^2(1-u)^2 M^G(\mu^2) B(\xi,W)\, ,
\\[4mm]  \displaystyle
\Phi^Q (u, \xi, W^2)=
-30 u(1-u)\ (2u-1)\frac1{N_f} M^Q(\mu^2) B(\xi,W) \, .
\ea 
\ee

As at present a little is known about $2 \pi$ DA's, the main motivation beyond
the model (\ref{DAmodel}) is that it incorporates all constraints arising from
crossing symmetry and soft pion theorems and has a simple form which makes its
treatment in numerical calculations easy. 
Note that dominance of the lowest conformal wave seems to be
phenomenologically justified in the case of the single pion DA.

This model differs from the asymptotic DA (\re{DAasW}) only in
values of momentum fractions $M^G(\mu^2)$ and $M^Q(\mu^2)$.
Their scale-dependence 
is given by
\be{Mevol}
M^Q(Q^2)=M^Q_{as}\left( 1+
L(Q^2)R(\mu^2)\right ),
\quad  R(\mu^2)= \frac{M^Q(\mu^2)-M^Q_{as}}{M^Q_{as}}\, , 
\ee
where $L$ is the usual evolution factor:  
\be{L}
L(Q^2)= \left( \frac{ \alpha_S(Q^2)}{\alpha_S(\mu^2)}  \right)^
{\gamma_{+}/b_0}, \quad \gamma_{+}=\frac23(N_f+4C_F), 
\, b_0=\frac{11}{3}-\frac23 N_f.
\ee
Obviously, that  $ M^G(Q^2)=1-M^Q(Q^2)$.

Explicit expression for $B(\xi,W)$ can be obtained using the
Watson theorem \ci{Polyakov}. In calculations considered here
$B(\xi,W)$ enters as a $Q^2$-independent factor and therefore its explicit
functional form is not important for considerations of power-suppressed
corrections to the amplitude.
To remove this factor from numerical calculations
  we will consider the $Q^2$-dependence of the ratio
$T^{\gamma\pi\pi}_0(Q^2,\xi,W^2 )/T_0^{as}$, where
\be{Tas}
T_0^{as}=T_0^{\gamma\pi\pi}(Q^2=\infty,\xi,W^2)=\sum_q e_q^2  
\int_0^1 du \frac{2u\Phi^Q_{as}(u,\xi,W^2)}{1-u}=
-\frac{15}{14}\sum_q e_q^2B(\xi, W^2) \, .
\ee
is asymptotic value of the amplitude. 

 In Fig.~\re{T0NLO} we compare
this ratio for the NLO amplitude $T^{pert}_0(Q^2,\xi,W^2 )$
for different models of $2\pi$ distribution amplitudes:
asymptotic (\re{DAasW}) and its minimal extension (\re{DAmodel}).  
Note that there is an ambiguity due to the scale-dependence of the NLO
amplitude. In order to estimate this uncertainty 
we make the following choice for the scale $\mu$:
\be{prtscl} 
\mu^2= \kappa Q^2+M^2\, ,
\ee  
with $M^2 = 1$ GeV$^2$ and a parameter $\kappa$ which will be varied
in the interval $1/5 \le \kappa \le 1$. Such a choice is motivated by
an observation \ci{Braun99} that a typical virtuality of a propagator
in a perturbative, exclusive amplitude is given by a weighted sum of
the hard scale $Q^2$ and the infrared cut-off. In the LCSR approach
the latter role is played by the Borel mass $M^2$, see the next
section.  Evaluating the NLO amplitude (\ref{NLOLz0}) we have
neglected numerically small NLO corrections \ci{Bel98,Rad86,DMull95}
to evolution of distribution amplitudes in the $\overline{MS}$
scheme.\\

One finds that in the model based on asymptotic distribution amplitudes the
NLO correction is numerically much larger than in the second model, where
amplitudes have the asymptotic form but different, scale-dependent
normalization. It can be understood by observing that  the gluon
distribution amplitude gives the main contribution to the NLO
correction. At $Q^2$ of order of few GeV$^2$ its normalization in
(\re{DAmodel}), based on the momentum fraction carried by gluons in a pion
according to the GRV parametrization \ci{GRV} is much smaller than the
asymptotic one. Strong sensitivity of the NLO corrections to gluon
distribution amplitude is an interesting feature of the process considered
here. As it follows, we have to revise one of our conclusions from \ci{KMP},
about the size of the NLO correction. Contrary to our previous claim, its
magnitude turns out to be rather model dependent, and therefore it is
difficult to make a trustworthy prediction unless distribution amplitudes
which enter the NLO correction, in particular the gluon one, are sufficiently
constrained.

\section{\normalsize \bf Light-cone sum rules method: the LO approximation}
 
In this section we discuss the Light Cone Sum Rule (LCSR) for the amplitude
$T^{\gamma\pi\pi}_0$. Our procedure closely follows investigation of the
photon-pion transition form-factor $F_{\gamma\pi}(Q^2)$ in Ref. \ci{Khod}.

The first step to obtain a LCSR for the amplitude with one real photon
$T^{\gamma\pi\pi}_0$ is to consider an amplitude $T^{\gamma^*\pi\pi}_0$ 
where both photons are off-shell and have large virtualities:
\be{Proc}
\gamma^*(q)+\gamma^*(q')\rightarrow 2\pi (P,k), Q^2, Q^{\prime \, 2} \gg
\Lambda_{\rm QCD}^2 
\ee
and find its dispersion representation in the variable $q^{\prime \, 2}$.

In the general kinematics momenta $q$ and $q^\prime$ can be
represented in terms of vectors $p$ and $z$ as
\be{lcexpansion1}
\ba{l}
\displaystyle
q = - \frac{Q^2}{2\sigma (p\cdot z)}\, p + \sigma\, z, \, q^2 = - Q^2\, 
\hspace*{8mm}
q^\prime = - \frac{Q^{\prime \, 2}}{2\alpha (p\cdot z)}\, p + \alpha \, z, 
q^{\prime \, 2} = - Q^{\prime \, 2} \, .
\ea
\ee
Coefficients $\alpha$ and $\sigma$ are functions of
kinematical invariants. Using momentum conservation one finds
\be{kinematics}
\ba{l}
\alpha = \frac{W^2}{2 (p\cdot z)} - \sigma, 
\\[4mm]\displaystyle
\sigma = \frac{1}{2 (p\cdot z)}(q \cdot (q - q^\prime) - \sqrt{X}),
\hspace*{8mm} X = (q\cdot q^\prime)^2 - q^2 q^{\prime 2} \, .
\ea
\ee
\\
With the help of the factorization theorem one can write $T^{\gamma^*\pi\pi}_0$
as a convolution of hard and soft blocks. Virtuality $q^{\prime \,
  2}$ enters the hard part only, where we neglect $W^2$ as compared with $Q^2$
and
$Q^{\prime \, 2}$. 

With $q^{\prime \, 2} \neq 0$ the two-photon amplitude admits naturally a
reacher Lorentz structure than the original amplitude with one photon
on-shell. In addition, by splitting $T^{\mu\nu}$ into Lorentz tensors and
invariant coefficient functions it is advantageous to avoid kinematical
constraints for the latter. Constraints imposed on coefficient functions
result in constraints on the form of their
dispersion representation.

To this end we rewrite the transverse metric tensor in terms
of momenta $q$ and $q^\prime$ \ci{Budnev}: 
\be{definitionR}
R^{\mu\nu}(q,q^\prime) = (-g^{\mu\nu})_T = -g^{\mu\nu} + \frac{1}{X}(q \cdot
q^\prime (q^\mu q^{\prime \, \nu} + q^\nu q^{\prime \, \mu}) - q^2 q^{\prime
  \, \mu} q^{\prime  \, \nu} - q^{\prime \, 2} q^\mu q^\nu),
\ee
with $X$ defined in (\re{kinematics}), and introduce a variable $\omega$
\be{definitions}
\ba{l}
\displaystyle
{\omega} = \frac{2P\cdot q}{q^2}=\frac{W^2}{Q^2}+\frac{q^2-q'^2}{q^2}
\simeq\frac{q^2-q'^2}{q^2} \, ,  
\ea
\ee
With these definitions it is convenient to introduce $T^{\gamma^*\pi\pi}_0$ as
\be{T0}
T^{\mu\nu}= i\frac{\omega^2}2 R^{\mu\nu}(q,q^\prime)
T^{\gamma^*\pi\pi}_0(Q,q',\xi,W)+ \dots
\ee
where ellipsis denotes other possible Lorentz structures.  Now, in the limit
$-q'^2\rightarrow 0$ ($\omega \rightarrow 1 $) $T^{\gamma^*\pi\pi}_0$, defined
here, goes smoothly into $T^{\gamma\pi\pi}_0$ given by (\re{NLOLz0}).  On the
other hand, factor
$\omega^2$ cancels singularity present in the tensor $R^{\mu\nu}$ as $q^2 \to
q^{\prime \, 2}, W^2 \to 0$ ($\omega \rightarrow 0 $) due to the $1/X$ term,
hence no constraints have to be imposed on $T^{\gamma^*\pi\pi}_0$.

To the LO accuracy and keeping $W^2 = 0$ in the hard block
the amplitude
$T^{\gamma^*\pi\pi}_0(Q,q',\xi,W)$ is the only one which appears in
$T^{\mu\nu}$ and one
obtains:
\be{LO}
\ba{l}\displaystyle
T^{\gamma^*\pi\pi}_0(Q,q',\xi,W)=
(\sum_q e_q^2 )\, 
 \int_0^1 dx\,
\Phi^Q(x,\xi,W^2)\frac{2x}{1-x\omega}
=
\\[4mm]\displaystyle
=
(\sum_q e_q^2 )\, 
 \int_0^1 dx\,
\Phi^Q(x,\xi,W^2)\frac{2x Q^2}{(1-x)Q^2-xq'^2}
\ea
\ee
Note that the above expression depends on $W$ through the 2-pion distribution
amplitude (\re{quarkDA}).

The LO result can be easily converted into a dispersion integral
over $q'^2$ with $s=\bar{x}Q^2/x$ being the mass of the intermediate state.
Using symmetry properties of the quark DA one finds
\be{DispQCD}
T^{\gamma^*\pi\pi}_0(Q,q',\xi,W)=
\int_0^{\infty}ds\frac{\rho_0(Q,s,\xi,W)}{s-q'^2}
\ee 
where
\be{RQCD}
\rho_0(Q,s,\xi,W)=2 \, (\sum_q e_q^2 )
\int_0^1 dx\,\delta(x-\frac{Q^2}{s+Q^2})x^2\Phi^Q(x,\xi,W^2) 
\ee

The next step is to rewrite the dispersion relation in the $q'^2$ channel
assuming that the spectral density can be approximated by contributions of
low-lying hadron states $\rho, \omega $ and a continuum of higher-mass states
with an effective threshold $s_0$:
\be{DispH}
T^{\gamma^*\pi\pi}_0(Q,q',\xi,W)= \sqrt{2}f_\rho
\frac{T^{\rho\pi\pi}(Q,\xi,W)}{m_\rho^2-q'^2}+
\int_{s_0}^{\infty}ds\frac{\rho^{cont}(Q,s,\xi,W)}{s-q'^2} \, .
\ee 
Here we have introduced the following notation for the matrix elements:
\be{mrho}
\ba{l}
\displaystyle
\langle \rho^0|J_\nu|0\rangle = \frac{1}{\sqrt{2}}f_\rho m_\rho\epsilon^*_\nu, 
\\[6mm]\displaystyle
\langle \pi\pi|J^\mu|\rho^0\rangle = \frac i{m_\rho}\frac{\omega^2}2
R^{\mu\nu}(q,q')\epsilon_\nu
T^{\rho\pi\pi}_0(Q,\xi,W)+ \dots 
\ea
\ee
$\epsilon^*_\nu,\epsilon_\nu $ are the polarization vectors of the $\rho$ meson.
To include  $\omega$-meson we adopt the   
following approximate relations
\be{approx}
m_\omega \simeq m_\rho,\,  3f_\omega\simeq f_\rho,\, 
T^{\omega\pi\pi}_0(Q,\xi,W)\simeq 3T^{\rho\pi\pi}_0(Q,\xi,W)
\ee
which follow from the quark content of $\rho$ and $\omega$ and from the 
isospin symmetry.

The right-hand-side (RHS) of equation (\ref{DispH}) involves two unknown
functions: the form-factor $T^{\rho\pi\pi}_0$ and the spectral density
$\rho^{cont}$.  Assuming quark-hadron duality one can estimate the continuum
spectral density as
\be{continum}
\rho^{cont}(Q,s,\xi,W)=\theta(s>s_0)\rho_0(Q,s,\xi,W)
\ee
where $\rho_0(Q,s,\xi,W)$ is the spectral density given in
(\ref{RQCD}).
By keeping $-q'^2$ large  and combining (\ref{continum}), (\ref{DispH}) and 
(\ref{LO}) one obtains a LCSR for the form-factor $T^{\rho\pi\pi}_0$:
\be{Trho2pi}
\sqrt{2}f_\rho\frac{T^{\rho\pi\pi}_0(Q,\xi,W)}{m_\rho^2-q'^2}=
\int_0^{s_0}ds\frac{\rho_0(Q,s,\xi,W)}{s-q'^2}
\ee
After perfoming Borel transformation in $-q'^2$ one finds ($M$ is the Borel
mass):
\be{Trho2piBorel}
\sqrt{2}f_\rho{T^{\rho\pi\pi}_0(Q,\xi,W)}=
\int_0^{s_0}ds{\rho_0(Q,s,\xi,W)}e^{(m_\rho^2-s)/M^2}
\ee
Finally, substituting (\ref{continum}) and (\ref{Trho2piBorel}) into
(\ref{DispH}) results in
\be{Disp}
T^{\gamma^*\pi\pi}_0(Q,q'^2,\xi,W)=\frac1{m_\rho^2-q'^2}
\int_0^{s_0}ds{\rho_0(Q,s,\xi,W)}e^{(m_\rho^2-s)/M^2}+
\int_{s_0}^{\infty}ds\frac{\rho_0(Q,s,\xi,W)}{s-q'^2}
\ee 
The above representation allows to
perform an analitical continuation to the  point $q'^2=0$. In this way
one arrives at a LCSR for the amplitude with one real photon:
\be{LOSumR}
T^{\gamma\pi\pi}_0(Q,\xi,W)=\frac1{m_\rho^2}
\int_0^{s_0}ds{\rho_0(Q,s,\xi,W)}e^{(m_\rho^2-s)/M^2}+
\int_{s_0}^{\infty}\frac{ds}{s}\rho_0(Q,s,\xi,W)
\ee    

It is convenient to rewrite this formula going back to an integral over 
fraction $x=Q^2/(s+Q^2)$. Introducing $x_0=Q^2/(s_0+Q^2)$ and 
$\bar x \equiv 1-x$ one finds
\be{LCSR}
T^{\gamma\pi\pi}_0(Q,\xi,W)&=&\frac{2Q^2}{m_\rho^2}(\sum_q e_q^2 )\,
\int_{x_0}^1dx {\Phi^Q (x, \xi, W^2)}\exp 
\left\{\frac{x m_\rho^2-\bar x Q^2}{xM^2}\right\}+
\\[4mm]\nonumber
&& +(\sum_q e_q^2 )\, \int_{0}^{x_0}dx\frac{2x}{1-x}\Phi^Q (x, \xi, W^2) \, .
\ee    

Consider now the limit $Q^2\rightarrow \infty$. In this limit
$x_0=1-s_0/Q^2+O(1/Q^4)$. The integration region in the first term
in the RHS (\re{LCSR}) shrinks 
to the point $x=1$ and one obtains
\be{1term}
\frac{2Q^2}{m_\rho^2}\int_{x_0}^1dx {\Phi^Q (x, \xi, W^2)}
\exp\left\{\frac{x m_\rho^2-\bar x Q^2}{xM^2}\right\}
=
\frac{2s_0^2}{Q^2 m_\rho^2}
\Phi^Q_x (1, \xi, W^2)\int_0^1 dx x e^{\frac{m_\rho^2-xs_0}{M^2}}+O(1/Q^4)
\ee
where $\Phi^Q_x (1, \xi, W^2)\equiv \frac d{dx}\Phi^Q (x, \xi, W)|_{x=1}$.

As it has been discussed at length in the literature, equation
(\re{1term}) can be interpreted as the so-called ``end-point''
contribution which arises from large transverse dinstances between two
photons in the hard block \ci{Braun99, BH}.  In general, the 2$\pi$
distribution amplitiude $\Phi^Q (x, \xi, W^2)$ depends also on the
factorization scale $\mu$ which separates large and short
distances. As $x\sim 1$ the quark virtuality (the magnitude of the
quark denominator in equation (\re{LO})) is of order of the Borel mass
$M$. As it follows, the two pion DA in (\re{1term}) should be
evaluated at a low normalization point, of order of $M$.

In Fig.\re{fig1}  we show an average value of the momentum fraction
$x$ in the integral (\re{LCSR}) calculated as a function of $Q^2$.
One observes that the mean value of $x$ in (\re{LCSR}) is indeed close to 1.

As $Q^2\rightarrow \infty$ the second term in the RHS of (\re{LCSR}) gives 
\be{2term}
\int_{0}^{x_0}dx\frac{2x}{1-x}\Phi^Q (x, \xi, W^2)=
\int_{0}^{1}dx\frac{2x}{1-x}\Phi^Q (x, \xi, W^2) +O(1/Q^2) \, .
\ee
It reproduces the leading order, leading twist factorization 
formula when $q'^2=0$ in (\re{LO}) and
provides correct asymptotics for very large $Q^2$. 
The power correction is suppressed as $1/Q^2$.

The LO sum rule (\re{LCSR}) results in an expression for the amplitude
$T^{\gamma\pi\pi}$ which includes contributions from both the ``end-point''
region $x \sim 1$, associated with large transverse distances, and from small
transverse distances where the $q {\bar q}$ pair is created by two photons in
a compact configuration. From the sum rule it follows that the ``end-point''
contribution is suppressed by $1/Q^2$ as compared with the LO factorization
result, i.e. it has the same order as factorizable higher twist corrections.
Despite formal power suppression, the ``end-point'' contribution can be
numerically important for realistic values of $Q^2$. Our numerical analysis
suggests that the sum rule (\re{LCSR}) can be applied for description of the
amplitude $T^{\gamma\pi\pi}$ starting from moderate momentum transfers 
$Q^2\ge 1$GeV$^2$.

\section{\normalsize \bf Radiative corrections} 
 
In principle, the sum rule (\re{LCSR}) can be improved in a twofold way, by
including the NLO $\alpha_s$ and higher-twist corrections to the spectral
density (\re{RQCD}). In this section we consider the NLO contribution.  Taking
into account higher-twists requires knowledge of the corresponding 2-pion
distribution amplitudes which are not known yet.

One-loop corrections for the real photon case have been considered in
\ci{KMP}. In current situation one should calculate the coefficient functions
in kinematics when both photons are vitual.  One should also keep in mind that
at the NLO diagrams with gluons enter the game.  As in \ci{KMP} one can
use crossing symmetry to derive coefficient functions from the corresponding
coefficient functions in the DVCS kinematics, as computed in
\ci{Belxx,JiOsb98,Manketal98}. The calculation is straightforward. The NLO
amplitude can be written in the standard form:
\be{T0-NLO}
\ba{l}
\displaystyle
T^{\gamma^*\pi\pi}_0(Q,q',\xi,W)=
\sum_q e_q^2 \, \left( 
 \int_0^1 dx\,
\Phi^Q(x,\xi,W^2) \left[C^0_q(x,\omega)+
{\textstyle \frac{\alf_S(\mu^2)}{4\pi}}
C^1_q(x,\omega,\mu)\right]- \right.
\\[6mm]\displaystyle \mskip150mu \left.
-\int_0^1 dx\,
\Phi^G(x,\xi,W^2)\left[
{\textstyle \frac{\alf_S(\mu^2)}{4\pi}} C^1_g(z,\omega,\mu)
\right] \,\right) .
\ea
\ee
Coefficient functions $C^0_q, C^1_q$ and $C^1_g$ are collected in the
Appendix.

As in the LO case, virtuality $q^{\prime \, 2}$ enters only through the
variable $\omega$.  It is convenient to rewrite equation (\re{T0-NLO}) in the
form which resembles the structure of the LO answer (\re{LO}):
\be{rho}
T^{\gamma^*\pi\pi}_0(Q,q',\xi,W)=
 \, \int_0^1 dx\frac{x\rho(x,\xi,W^2)}{1-x\omega} 
\ee
with
\be{rhodef}
\rho(x,\xi,W^2)= \rho_0(x,\xi,W^2)+ 
\frac{\alf_S(\mu^2)}{4\pi} \rho_1(x,\xi,W^2) \, .
\ee
As in equation (\re{LO})
\be{rho0}
\rho_0(x,\xi,W^2)=2 (\sum e_q^2) \Phi^Q(x,\xi,W^2) \, .
\ee       
The NLO correction $\rho_1(x,\xi,W^2)$ can be obtained from
the corresponding coefficient functions (\re{T0-NLO}):
\be{rhoCrel}
\ba{l}\displaystyle
\sum e_q^2 \int_0^1 dx \left[\Phi^Q(x,\xi,W^2)
\frac1\pi\mbox{Im}_{q'^2}C_q^1(x,\omega)-
 \Phi^G(x,\xi,W^2)\frac1\pi\mbox{Im}_{q'^2}C_g^1(x,\omega)\right] =
\\[6mm]\displaystyle
\mskip200mu 
= u^2\rho_1(u,\xi,W^2)\large|_{u=1/\omega}
\ea
\ee
where $\mbox{Im}_{q'^2}$ denotes the imaginary part with respect to the
variable $q^{\prime \, 2}$, considered here as a complex variable with
positive both real and (infinitesimal) imaginary parts.


With such a definition the structure of the NLO LCSR for the amplitude
$T^{\gamma\pi\pi}_0$ remains the same as the LO one:
\be{LCSRNLO}
T^{\gamma\pi\pi}_0(Q,\xi,W)=\frac{Q^2}{m_\rho^2}
\int_{x_0}^1dx \rho(x,\xi,W^2) \exp 
\left\{\frac{x m_\rho^2-\bar x Q^2}{xM^2}\right\}+
\int_{0}^{x_0}dx\frac{x}{1-x}\rho(x,\xi,W^2) \, .
\ee       
As in the case of the LO sum rule (\ref{LCSR}), as $Q^2 \to \infty$
equation (\ref{LCSRNLO}) reproduces the perturbative expansion of the
$T^{\gamma\pi\pi}_0$ amplitude, including the NLO corrections
\cite{KMP}.
The power correction is suppressed as
$1/Q^2$.

As an illustration, we quote now the form of $\rho_1(x,\xi,W^2)$ in the case
of the 'minimal model' (\re{DAmodel})
\be{minrho1}
\ba{l}\displaystyle
\rho_1(x,\xi,W^2)=\sum e_q^2 B(\xi,W^2)\left\{
\ln(\mu^2/Q^2)\rho_{10}(x,\mu^2)+\rho_{11}(x,\mu^2)
\right\}, 
\\[4mm]\displaystyle
\rho_{10}(x,\mu^2)= (-40)x(1-x)(2x-1)R(\mu^2),
\\[4mm]\displaystyle
\rho_{11}(x,\mu^2)= \rho_{as}(x)+ R(\mu^2)[ \rho_{as}(x)+D(x)] \, .
\ea
\ee
Where $R(\mu^2)$ is defined in (\re{Mevol}). 
Function $D(x)$ is a shorthand notation for:
\be{D}
D(x)=-\frac{10}{3}[2\bar x (1-6x) + x\bar x (2x-1)(31+12\ln(x/\bar x) )]
\ee
Taking for $M^Q$ the asymptotic value (\re{Masymp}) results in
$R(\mu^2=\infty)= 0$ and 
\be{rho1as}
\rho_1(x,\xi,W^2)=B(\xi,W^2)\rho_{as}(x)
\ee 
where 
\be{rhoas}
\rho_{as}(x)= \frac{-20C_F}{N_f+4C_F}x(1-x)(2x-1)(\pi^2-17-3\ln^2(x/\bar x))
\ee
In this case $\rho_1$ has no dependence on $\mu$.

Before evaluating the sum rule one has to provide an estimate of the
factorisation scale $\mu$. This is a standard problem in a calculation
based on the fixed-order perturbation theory. 
Note that the scale of
perturbative expansion $\mu^2 \sim Q^2$ is different from the
characteristic scale of soft, 'end-point' contributions $\mu^2 \sim
M^2$. As a consequence, the correct treatment of the sum rule requires
applying various normalization scales to various terms in
(\re{LCSRNLO}).  To avoid confusion, we propose the following
procedure. One separates the perturbative contribution and writes the
final formula in the form:
\be{Tsep}
T^{\gamma\pi\pi}_0(Q,\xi,W)=
T_0^{\rm {pert}}(Q,\xi,W)\, +\, T_0^{\rm {non-pert}}(Q,\xi,W)\, .
\ee 
$T_0^{pert}$ is the perturbative amplitude (\re{NLOLz0}), which can be
represented as:
\be{Tperp}
T_0^{\rm {pert}}(Q,\xi,W)\, =\, 
\int_{0}^{1}dx\frac{x}{1-x}\rho(x,\xi,W^2) \, .
\ee
The scale in this term is set by the hard photon virtuality $Q^2$. 
The second term should, as a matter of fact, be considered as the proper 
light-cone sum rule result:
\be{Tnonp}
T_0^{\rm {non-pert}}(Q,\xi,W)\, =\,
\int_{x_0}^1dx \rho(x,\xi,W^2)
\left[ 
\frac{Q^2}{m_\rho^2}
\exp \left\{ \frac{x m_\rho^2-\bar x Q^2}{xM^2}\right\}
- \frac{x}{1-x}
\right]
\ee
Here the integration region is restricted to $x_0 \le x \le 1$. As 
discussed in the previous section, this term should be evaluated at a 
{\it low} normalization point $\sim M^2$.  

Of course, when evaluated at the same normalization scale, sum of
(\re{Tperp}) and (\re{Tnonp}) reproduces the original sum rule
(\re{LCSRNLO}). \\

\section{\normalsize \bf Numerical results}

In the subsequent numerical analysis the following input has been used:
the threshold parameter $s_0=1.5$ GeV$^2$ has been taken from the
two-point sum rule \ci{SVZ}.  This sum rule is reliable for the
corresponding Borel parameter $M^2_{2pt}=0.5-0.8$ GeV$^2$. In the
light-cone sum rules it should be larger to compensate for the fact
that the effective expansion parameter is given by the inverse power
of $xM^2$. In this case a typical choice is $M^2\sim M^2_{2pt}/\langle
x\rangle$ \ci{Khod, Braun99}.  We assume $0.6 \le M^2 \le 1.2$ GeV$^2$
as a reasonable interval. In addition we have checked that changing
$s_0$ by $\pm 0.2$ GeV$^2$ does not produce any sizeable effect.

We remind the reader that in the present investigation we have
neglected higher twist contributions to the sum rule for
$T_0^{\gamma^*\pi\pi}$. They are, as usual, suppressed by additional
powers of the Borel parameter. To obtain their contribution one should
know two-pion distribution amplitudes of higher twists which have not
yet been studied.  In the case of photon-pion transition form-factor
\ci{Khod}, the contribution of such terms to the non-perturbative
(power suppressed) part is about $30-35\%$ of the leading twist
contribution for the low $Q^2\sim 1-2$ GeV$^2$. Qualitatively, the
same picture should hold also in the present case.

All numerical results have been obtained with the model (\re{DAmodel}) for
$2\pi$ distribution amplitude. We have used the quark momentum fraction 
$M^Q(0.6\, {\rm GeV^2})=0.63$ taken from the GRV parametrization \ci{GRV}. 
We have kept $N_f=4$ which results in $M^Q_{as}=0.43$. 

In all calculations we have used $\Lambda_{QCD}=0.204$ GeV and 
one-loop running coupling  $\alpha_S(1 \, {\rm GeV}^2)=0.47$.

In the numerical analysis one has to specify the factorization scale.
Since after subtraction of the perturbative amplitude $T_0^{\rm
{non-pert}}(Q,\xi,W)$ is dominated by the ``end-point'', soft
contributions, we apply here a fixed, low scale $\mu^2\sim M^2$.  

To estimate the ambiguity due to the scale dependence in $T_0^{\rm
{pert}}(Q,\xi,W)$ we parametrize $\mu^2$ according to:
\be{ptscale}
\mu^2=\kappa Q^2+M^2 \, ,
\ee
with a parameter $\kappa$ which varies in the interval $ 1/5 \le \kappa \le 1$.
 
We consider values of $Q^2$ in the physically interesting interval $1 \,
{\mbox {GeV}}^2 \le Q^2 \le 10 \, {\mbox {GeV}}^2$.
  
To cancel the influence of the
overall factor $B(\xi, W^2)$, in the following we always plot the ratio of
$T_0^{\gamma\pi\pi}(Q^2,\xi,W^2)$ to its asymptotic value at $Q^2=\infty$
(\re{Tas}).\\

The Borel parameter dependence of the ratio $T^{\rm
{non-pert}}_0/T_0^{\rm as}$ is shown in Fig.~\re{Borel}. We find that
the $M^2$ dependence is sufficiently flat to justify the use of LCSR,
although for lower values of $Q^2$, where power corrections are
increasingly important, dependence on the Borel parameter becomes
stronger.  Assuming that the uncertainty arising from the Borel
parameter dependence should not exceed 30 \% for the $T_0^{\rm
{non-pert}}$, we estimate that the sum rule (\re{LCSR}) provides a
reasonable description of $T^{\gamma\pi\pi}_0$ starting from $Q^2 \ge
1$ GeV$^2$.

In Fig.~\re{fig4} we show the ratio $T_0^{\rm
{non-pert}}(Q^2)/T_0^{\rm as}$ obtained from the light-cone sum rule
(\re{Tnonp}) for different values of the Borel parameter $M^2$. One
observes that in the whole interval of $Q^2$ considered here the sum
rule calculation is rather stable with respect to variation of the
Borel parameter within a reasonable interval. In accordance with
expectations, the non-perturbative correction becomes smaller with
increasing $Q^2$. By fitting a simple formula we found that for the
value of the Borel parameter in the middle of the interval, $M^2 =
0.9$ GeV$^2$, and in the $Q^2$ region $1
\le Q^2 \le 10\, {\mbox {GeV}}^2$, $T^{\rm {non-pert}}_0/T_0^{\rm as}$ can
be parameterized as
\be{Tfit}
T^{\rm {non-pert}}_0(Q^2)/{T_0^{\rm as}}=
\frac{-1.5+0.05Q^2+0.015 Q^4}{1+Q^2+0.3Q^4} \, ,
\ee
with $Q^2$ in units of GeV$^2$, see the solid
line in Fig.~\re{fig4}.

In Fig.~\re{resultfig} we present perturbative and non-perturbative
contributions to the ratio $T^{\gamma\pi\pi}_0/T_0^{\rm as}$ as a
function of $Q^2$. $T_0^{\rm {non-pert}}$ is calculated with
$M^2=0.9\, {\mbox {GeV}}^2$. The non-perturbative corrections is
numerically significant only in the region $Q^2 \le 4$ GeV$^2$. For
higher values of $Q^2$ the amplitude is dominated by the NLO
leading-twist contribution.

\section{\bf Summary and conclusions}

The main result of this paper is the numerical estimate of
power-suppressed correction to the leading-twist helicity-conserving
amplitude of the process $\gamma^* \gamma \to \pi \pi$. Light-cone sum
rules technique allows to circumvent difficulties due to
non-factorizability of the power-suppressed terms. Although formally
our analysis is not complete, as we have neglected contribution of
higher twist operators to the amplitude of two-pion production in a
collision of two virtual photons, we believe that the general picture
is reliable, at least qualitatively. Power corrections are
increasingly important with decreasing $Q^2$ for $Q^2 \le 4$ GeV$^2$,
and become about 50\% of the leading-twist amplitude at $Q^2 = 1$
GeV$^2$.

Our final result for the helicity-conserving $\gamma^* \gamma \to \pi \pi$
amplitude is shown in Fig.~\re{finfig}. The grey bound indicates uncertainty
due to the linearly combined Borel parameter and the factorization scale
variations. 
One finds that starting from $Q^2$ around 4 GeV$^2$ the twist-2
contribution approximately saturates the amplitude. This observation suggests
that the cross-section of the process $\gamma^* \gamma \to \pi \pi$ can be 
accurately predicted in QCD for given models of $2\pi$ distribution amplitudes.

Assuming dominance of the lowest conformal wave, we have found that helicity
conserving amplitude is very sensitive to the normalization of gluonic $2\pi$
distribution amplitude. This observation, combined with crossing,  makes it
plausible to use $\gamma^* \gamma \to \pi \pi$ to constrain momentum fraction
carried by gluons in a pion.

\bigskip

{\bf Acknowledgments}: We gratefully acknowledge discussions with
V. Braun, M. Polyakov, A.~Belitsky and O.~Teryaev. We are grateful to
M.~Diehl for his valuable comments. This work was supported in part by
KBN grant 2~P03B~011~19 and DFG, project No.~920585.

\section{\normalsize \bf APPENDIX:  Coefficient functions  }
\setcounter{equation}{0}
\label{app:a}
\renewcommand{\theequation}{A.\arabic{equation}}
\setcounter{table}{0}
\renewcommand{\thetable}{\Alph{table}}

To the NLO accuracy the amplitude $T^{\gamma^*\pi\pi}_0$ can be represented as
a standard
convolution of $2\pi$ distribution amplitudes with the hard scattering
coefficients:
\be{NLOamplitude}
\ba{l}
\displaystyle
T^{\gamma^*\pi\pi}_0(Q,q',\xi,W)=
\sum_q e_q^2 \, \left( 
 \int_0^1 dx\,
\Phi^Q(x,\xi,W^2) \left[C^0_q(x,\omega)+
{\textstyle \frac{\alf_S(\mu^2)}{4\pi}}
C^1_q(x,\omega,\mu^2/Q^2)\right]- \right.
\\[6mm]\displaystyle \mskip150mu \left.
-\int_0^1 dx\,
\Phi^G(x,\xi,W^2)\left[
{\textstyle \frac{\alf_S(\mu^2)}{4\pi}} C^1_g(z,\omega,\mu^2/Q^2)
\right] \,\right) \, ,
\ea
\ee
where $\omega$ has been defined in (\re{definitions}).
The coefficient functions in the $\overline{MS}$ scheme read
\be{C1q}
C^0_q(x,\omega)&=&\frac{2x}{1-x\omega },
\\[6mm]\nonumber
 C^1_q(x,\omega,\mu^2/Q^2)&=& C_F\left[ 
\ln (\mu^2/Q^2)[C^{10}_q(x,\omega)-C^{10}_q(1-x,\omega)]+ 
C^{11}_q(x,\omega)-C^{11}_q(1-x,\omega)\right],
\\[6mm]\nonumber
C^{10}_q(x,\omega)&=& -\frac 3{(1-x\omega)\omega}-
\ln (1-\omega)\frac{2(1-\omega)}{\bar x \omega^2}
\left(\frac{1}{1-x\omega}-\frac{1}{\omega}\right)+
\\[6mm]
&&{}+2\, \frac{(1-\omega)}{1- x \omega}\frac{\ln (1- x \omega)}{\bar x \omega^2}+
 2\frac{\ln (1- x \omega)}{x \bar x \omega^2}
\left[
\frac{1- x \omega}{\omega}-\frac{\bar x }{1- x \omega} 
\right] \, ,
\\[6mm]\nonumber
C^{11}_q(x,\omega)&=& \frac1{(1-x\omega)\omega}\left[
-9+\frac2{x\omega}\ln^2(1-x\omega)-\frac3{x\omega}\ln(1-x\omega)+
\frac3{x\omega}\ln(1- \omega)-\right.
\\[6mm]\nonumber
&&{}
\left.
-\frac1{x\omega}\ln^2(1-\omega)  
\right]+
\frac{3}{\bar x \omega^2}\ln (1-x\omega)-\frac{3}{x \bar x \omega^2}\ln (1-\omega)+
\\[6mm]\nonumber
&&{}+(x-1-1/\omega)\frac{\ln^2 (1-x\omega)}{x \bar x \omega^2}+ 
(1-x+x/\omega)\frac{\ln^2 (1-\omega)}{x \bar x \omega^2}, 
\ee
\vskip 0.5cm
\be{CG}
C^1_g(x,\omega,\mu^2/Q^2)&=& \ln (\mu^2/Q^2)C^{10}_g(x,\omega)+ C^{11}_g(x,\omega),
\ee
\be{C10g}
C^{10}_g(x,\omega)&=& \frac{(-2)}{w^4(x\bar x)^2}\left\{
(1-\omega +[1-x\omega]^2)\ln (1-x\omega)+\right.
 \\[6mm]\nonumber 
&& \left. \frac{}{}
+(1-\omega +[1-\bar x\omega]^2)\ln (1-\bar x\omega)-
(2-\omega x^2-\omega {\bar x}^2)(1-\omega)\ln (1-\omega)\right\}, 
\ee

\be{C11g}
C^{11}_g(x,\omega)&=& 
\frac{1}{w^4(x\bar x)^2}\left\{a_1(x,\omega)\ln (1-\omega)+
a_2(x,\omega)\ln ^2(1-\omega)+\right.
\\[6mm]\nonumber
&& c_1(x,\omega)\ln (1-x\omega)+c_1(\bar x,\omega)\ln (1-\bar x\omega)+
\\[6mm]\nonumber 
&& \left. \frac{}{}
c_2(x,\omega)\ln^2 (1-x\omega)+ c_2(\bar x,\omega)\ln^2 (1-\bar x\omega)
\right\},
\\[6mm]\nonumber 
a_1(x,\omega)&=& 8+4\omega(x-3-x^2)+4\omega^2(1-x+x^2),
\\[6mm]\nonumber 
a_2(x,\omega)&=&-2+\omega(3-2x+2x^2)-\omega^2(1-2x+2x^2),
\\[6mm]\nonumber 
c_1(x,\omega)&=&-8+4\omega(1+2x)-2\omega^2x(1+x),
\\[6mm]\nonumber 
c_2(x,\omega)&=&2-\omega(1+2x)+\omega^2x^2 \, .
\ee
Here we used the shorthand notation $\bar x\equiv 1-x$. Note that
physical amplitude does not have a singularity (pole) when $\omega\rightarrow
0$  and therefore
all coefficient functions must be well defined in this limit: 
\be{limw0}
\ba{l}\displaystyle
C^1_q(x,\omega,\mu^2/Q^2)= 
C_F\left[ \ln (\mu^2/Q^2)\frac83(2x-1)+ (1-2x)\right]+O(\omega),
\\[4mm]\displaystyle
C^1_g(x,\omega,\mu^2/Q^2)= \ln (\mu^2/Q^2)\frac43+\frac73+O(\omega).
\ea
\ee

\newpage

\input pic1.tex

\input pic2.tex

\newpage

\input pic3.tex

\input pic4.tex

\newpage

\input pic5.tex

\input pic6.tex

\end{document}

%% file: pic1.tex
\begin{figure}[htb]
\begin{center}
\hspace{0cm}
\insertfig{12.5}{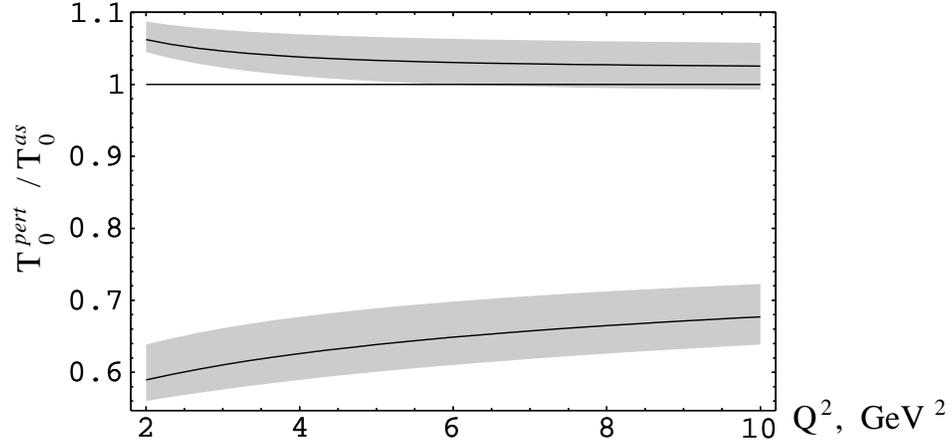}
\end{center}
\vspace{-0.5cm}
\caption[dummy]{
\small 
NLO results for the $T_0^{pert}(Q^2)/T_0^{\rm as}$ as a function of
$Q^2$. The upper plot corresponds to the mimimal model with
$M^Q(1\, {\mbox {GeV}}^2)=0.60$ and the lower one to the asymptotic choice
$M^Q = 0.43$.  The scale is taken as $\mu^2 = \kappa Q^2+M^2$,  
$M^2=1$ GeV$^2$. Gray bands show variation of the NLO prediction when 
$\kappa$ is varied between $1/5$ and $1$.  Solid lines correspond to 
$\kappa=2/5$.
\label{T0NLO}
}
\end{figure}

%% file: pic2.tex
\begin{figure}[htb]
\begin{center}
\hspace{0cm}
\insertfig{12.5}{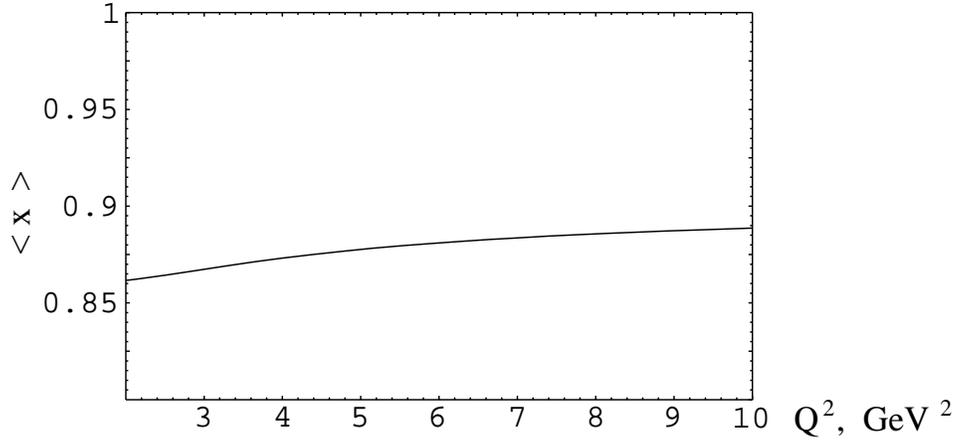}
\end{center}
\vspace{-0.5cm}
\caption{\small  
   Average momentum fraction $x$ as a function
$Q^2$. We take  $M^2=0.9$ GeV$^2 $ and $s_0=1.5$ GeV$^2$,
see explanation in the main text.  
\label{fig1}}
\end{figure}

%% file: pic3.tex
\begin{figure}[htb]
\begin{center}
\hspace{0cm}
\insertfig{12.5}{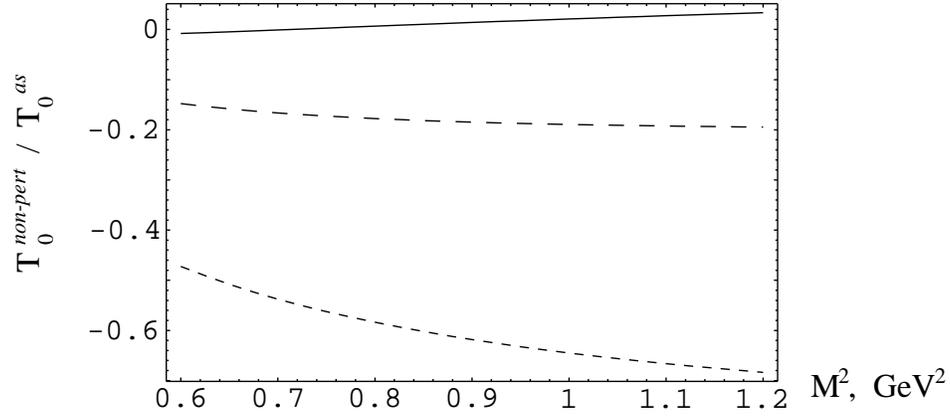}
\end{center}
\vspace{-0.5cm}
\caption[dummy]{\small 
Ratio $T^{\rm {non-pert}}_0/T_0^{as}$ as a 
function of $M^2$ for $Q^2=1\, {\mbox{GeV}}^2$ (short dashed), 
$Q^2=3\, {\mbox {GeV}}^2$ (long dashed) 
and $Q^2=10\, {\mbox {GeV}}^2$ (solid line).
\label{Borel}}
\end{figure}

%% file: pic4.tex
\begin{figure}[htb]
\begin{center}
\hspace{0cm}
\insertfig{12.5}{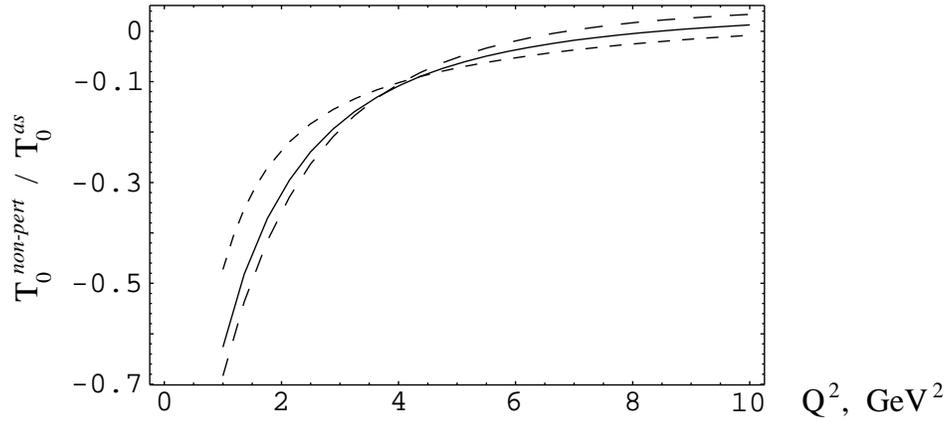}
\end{center}
\vspace{-0.5cm}
\caption[dummy]{\small Ratio $T_0^{\rm {non-pert}}/T_0^{\rm as}$ as a 
function of $Q^2$. Short- and long-dashed lines correspond to
$M^2=0.6$ and $1.2\, {\mbox {GeV}}^2$, with $\mu^2 = M^2$, 
respectively. Solid line ($M^2=0.9$ GeV$^2$) represents parametrization 
(\ref{Tfit}).
\label{fig4}}
\end{figure}

%% file: pic5.tex
\begin{figure}[htb]
\begin{center}
\hspace{0cm}
\insertfig{12.5}{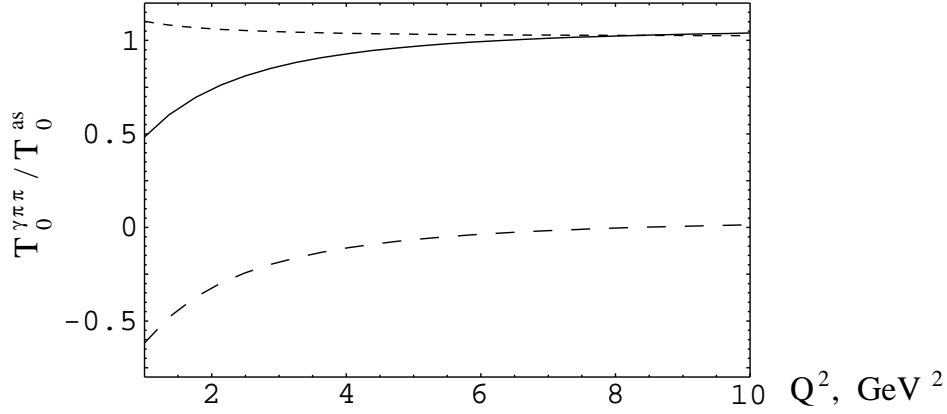}
\end{center}
\vspace{-0.5cm}
\caption[dummy]{\small 
The ratio $T_0^{\gamma\pi\pi}/T_0^{\rm as}$ as a function of $Q^2$
(solid line). Short-dashed line represents $T_0^{\rm pert}/T_0^{\rm
as}$ with $\mu^2=2/5\, Q^2+1 \, {\mbox {GeV}}^2$.  Long-dashed line
represents $T_0^{\rm non-pert}/T_0^{\rm as}$ with $\mu^2=M^2=0.9 \,
{\mbox {GeV}}^2$. Solid line is the sum of both.
\label{resultfig}}
\end{figure}

%% file: pic6.tex
\begin{figure}[htb]
\begin{center}
\hspace{0cm}
\insertfig{12.5}{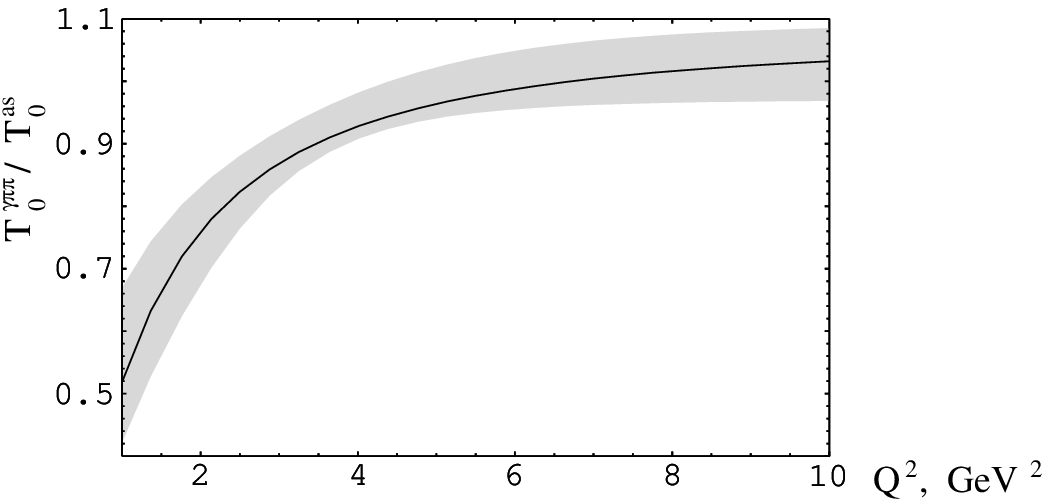}
\end{center}
\vspace{-0.5cm}
\caption[dummy]{\small The LCSR prediction for the ratio
$T_0^{\gamma\pi\pi}/T_0^{\rm as}$ as a function of $Q^2$. The grey
band shows the sensitivity of our result to variation of the Borel
parameter within $0.6 \le M^2 \le 1.2$ GeV$^2$ and factorization scale
according to formula (\ref{ptscale}).
\label{finfig}}
\end{figure}